\documentstyle[11pt]{article}

{\catcode `\@=11 \global\let\AddToReset=\@addtoreset}
\AddToReset{equation}{section} 

\newcommand{\leftineq}{\ll}
\newcommand{\rightineq}{\gg}

\newcommand{\ga}{\gamma}

\newcommand{\rh}{\scri}

\newcommand{\ph}{\phi}

\newcommand{\Ga}{\Gamma}

\newcommand{\Si}{\Sigma}

\newcommand{\commentout}[1]{}  
 
 \newcommand{\be}{\begin{equation}}

\newcommand{\ee}{\end{equation}} \newcommand{\bea}{\begin{eqnarray}}
\newcommand{\eea}{\end{eqnarray}}
\newcommand{\beaa}{\begin{eqnarray*}}
\newcommand{\eeaa}{\end{eqnarray*}} 

 \def\scri{\hbox{${\cal J}$\kern -.645em {\raise
      .57ex\hbox{$\scriptscriptstyle (\ $}}}}

\newcommand{\Bbb}{\bf}
\newcommand{\R}{{\Bbb R}}

\newcommand{\doc}{\leftineq\rh\rightineq}
\newcommand{\pdoc}{\partial\doc}

\newcommand{\W}{{\cal W}}
\newcommand{\B}{{\cal B}}
\newcommand{\ext}{{\mbox{\scriptsize ext}}}

     \newtheorem{Theorem}   {Theorem}   [section]
     
     \newtheorem{Lemma}     [Theorem]   {Lemma}
     \newtheorem{Proposition} [Theorem] {Proposition}
\begin{document}
\title{On the topology of stationary black holes} \author{Piotr T.\
  Chru\'sciel\thanks{Alexander von Humboldt fellow.  On leave of
    absence from the Institute of Mathematics, Polish Academy of
    Sciences, Warsaw.  Supported in part by a Polish Research Council
    grant KBN 2 P302 095 06 and by the Federal
    Ministry of Science and Research, Austria.
    Current address: D\'epartement de Math\'ematiques, Facult\'e des
    Sciences, Parc de Grandmont, F 37200 Tours, France. {\em e--mail}:
chrusciel@univ-tours.fr
    }~{}$^\ddagger$  \\Max Planck Institut f\"ur Astrophysik\\ Karl
    Schwarzschild Strasse 1\\ D 85740 Garching bei M\"unchen\\ \\
    Robert M.\ Wald\thanks{{\em e--mail}:
    rmwa@midway.uchicago.edu}~{}\thanks{Supported in part by NSF
    grant PHY--9220644 to University of Chicago.}
\\ Enrico Fermi
Institute and Department of Physics\\University of Chicago\\
5640 S. Ellis Ave., Chicago, IL 60637.}

\date{}

\maketitle

\begin{abstract}
We prove that the domain of outer communication of a stationary,
globally hyperbolic spacetime
satisfying the null energy condition must be simply connected. Under
suitable additional hypotheses,
this implies, in particular, that each connected component of a
cross-section of the event horizon of a stationary black hole must
have spherical topology.  \end{abstract}

\section{Introduction}

The theory of the uniqueness of stationary black holes in classical general
relativity intertwines the global techniques of differential geometry
with those of the
theory of geometric partial differential equations. In spite of considerable
progress in the understanding of
the issues involved, several open questions in that theory still
remain ({\em cf.\ e.g.\/} \cite{JaVancouver} for a
recent\footnote{Some of the questions raised in \cite{JaVancouver}
have  been settled in
\cite{HeuslerRNuniqueness,WaldRacz2,JaBeig}.} review). As
has been recently stressed by Galloway \cite{Ga}, one of those is the
expected spherical topology of connected components of spacelike
sections of event horizons. Recall that such a claim has been made in
\cite{Ha1,HE}, but, as discussed in detail in \cite{Ga}, the arguments
given there do not seem to exclude the possibility of toroidal
topology, except perhaps when analyticity up--to--and--including the
event horizon is assumed ({\em cf.}, however, \cite{Gannon} for a result
in the electrovacuum case with non--vanishing charge).
While this analyticity seems a plausible
property in retrospect, as a potential
consequence of the uniqueness theorems, no {\em a priori} reasons for
analyticity have been given so far. The object of this letter is to
point out that toroidal topology of stationary black holes -- as well
as all other non-spherical topologies -- can be excluded as a simple
consequence of the ``topological censorship theorem'' of Friedman,
Schleich and Witt \cite{FriedmanSchleichWitt}, when a suitable energy
condition is imposed. Moreover the differentiability conditions on the
event horizon implicitly assumed in \cite{HE} are not needed in our
argument to exclude the toroidal, as well as the higher genus
topologies.

We shall present the detailed statement of our Theorem in Section
\ref{main} below. Before doing that, let us point out that
some related results have been proved previously by Gannon \cite{Gannon},
by Galloway \cite{Ga,Gallowaybodies} ({\em cf.\/} also
\cite{GallowayFrankel}) and by Masood--ul--Alam
\cite{Masoodtopology} ({\em cf.\/} also \cite{BrillLindblom})
under various supplementary hypotheses. The
proof below arose as a byproduct of an attempt to gain insight
into the topology of black holes using the topological
censorship theorem of Galloway \cite{Ga}. A  related
application of the Friedman--Schleich--Witt topological censorship theorem
can be found in
\cite{Ted}.

\section{The theorem} \label{main}
We begin by arguing that it suffices to consider the case where
the spacetime has a single asymptotically flat region.
To make things precise,
let $(M,g_{ab})$ be a globally hyperbolic space-time
with Cauchy surface $\Si$ and with a one parameter
group of isometries, $\ph_t$, generated by Killing vector field $X^a$. $\Si$
will be assumed at first to have a (possibly infinite) number of asymptotic
regions $\Si_i$, in which $X^a$ is timelike and tends asymptotically
to a non-zero
multiple of the unit normal to $\Si$ as the distance away from some
fixed point $p\in \Si$ tends to infinity. Here the notions of
asymptotic flatness and of stationarity are used in the sense of
Definitions 2.1 and 2.4 of \cite{ChW}; we emphasize that $X^a$ is not
required to be globally timelike. We shall moreover assume that
the orbits of $X^a$ are complete on $M$. Let us mention that this last
hypothesis can be derived as a consequence of field equations and of
appropriate hypotheses on $\Si$, if field equations are assumed, {\em cf.\
e.g.\/} \cite{Chorbits} and \cite[Proposition 3.1]{ChW}.

Consider an asymptotically flat three-end $\Si_i$, and
let $\B_i$ and $\W_i$
be the black-- and white--hole regions with respect to $\Si_i$ as defined
in \cite{ChW}. Consider the domain of outer communication
$\leftineq\rh_i\rightineq $ defined as
$$
\leftineq\rh_i\rightineq =M\backslash\{\B_i\cup \W_i\}.
$$
The following result follows immediately from what has been said in
\cite{FriedmanSchleichWitt}:

\begin{Proposition} \label{P1} Under
the conditions above,
suppose moreover that the null energy condition holds
\begin{equation}
R_{ab}Y^aY^b\geq 0 \ \mbox{ for all null }\  Y^a\,. \label{NEC}
\end{equation}
Then
$$
\leftineq\rh_i\rightineq \cap \,J^\pm(\leftineq\rh_j\rightineq
)=\emptyset\  \mbox{ for } \ i\neq j\,.
$$
\end{Proposition}
In other words, the domain of outer communication associated to the
asymptotic three-end $\Si_i$ is causally separated from those
associated to the remaining asymptotic regions. Alternatively, when
analysing globally hyperbolic domains of outer communication in which
(\ref{NEC}) holds one can without loss of generality assume that the
relevant Cauchy surface has only one asymptotically flat
 region, as we desired to show.
Let us also mention that a somewhat similar result
has been proved in \cite[Lemmas 3.2 and 3.3]{ChW}. There it is assumed
that the time orientation of the Killing vector in the $i$'th end is
opposite to that in the $j$'th end, but no energy conditions are assumed.

In the following, we shall assume that $(M,g_{ab})$ contains a single
asymptotically flat region whose domain of outer communications will
be denoted by $\doc$.  We no longer need assume that $M$ is globally
hyperbolic, but we require global hyperbolicity of $\doc$ (which
automatically holds if $M$ is globally hyperbolic).  Let ${\cal H}
\equiv \partial I^-(\doc)$ denote the future event horizon of the
black holes of $M$. For some of our
results, we shall assume, in addition, that there exists an
achronal, asymptotically flat slice, ${\cal S}$, of $\doc$ whose
boundary in $M$ intersects
each null generator of ${\cal H}$ at precisely one point. In particular,
this implies that the topology of ${\cal H}$ is
$\R\times K$, where $K = \overline{{\cal S}} \cap {\cal H}$.
(We do not assume
that $K$ is connected.) Note that these hypotheses encompass, in
particular the case where no black hole is present (in which case $K$
is empty), the case of ``spacetimes of class (b)" as defined in
\cite{ChW}, as well as ``extreme black holes" (such as the
Papapetrou-Majumdar solution) not treated in \cite{ChW}.

Let $p$ be any point in the asymptotically flat region of $\doc$,
so that,
in particular, $X^a$ is timelike at $p$. Define ${\cal C} = \partial I^+(p)$.
Then ${\cal C}$ automatically is an achronal, $C^{1-}$ hypersurface.
We write ${\cal C'} = \doc \cap {\cal C}$, ${\cal C}_\ext = {\cal C}\cap
M_\ext$, with 
$ M_\ext$ denoting that part of $M$ which is covered
by a single coordinate system in which  the metric is asymptotically flat
and time--independent,
with the Killing vector being timelike there.
The proof of our main theorem below will make use of the following lemma:

\begin{Lemma} \label{L1}: Let $(M,g_{ab})$ be a
stationary, asymptotically flat
spacetime containing a single
asymptotically flat region whose domain
of outer communications, $\doc$, is globally hyperbolic. Then
\begin{enumerate}
\item
Each Killing orbit in $\doc$ intersects ${\cal C}$ in precisely one point;
hence, in particular, $\doc$ has topology $\R\times {\cal C'}$
\item
Suppose that there exists an achronal, asymptotically flat slice,
${\cal S}$, of $\doc$, whose boundary in $M$ intersects
the event horizon, ${\cal H}$, of any black holes
in $M$ in a cross-section, $K$.
If $K$ is compact, then each null generator of
${\cal H}$ intersects ${\cal C}$ in precisely one point; hence, in particular,
$\partial {\cal C'}$ has topology $K$.
\end{enumerate}\end{Lemma}

{\bf Proof:} To prove the first statement, we note that, by the same
argument as used in the proof of lemma 3.1 of \cite{ChW}, for any
Killing orbit, $\alpha$, in $\doc$, we have $\doc \subset I^+(\alpha)$
and $\doc \subset I^-(\alpha)$. Consequently, any Killing orbit in
$\doc$ enters both $I^+(p)$ and $I^-(p)$, and, thus, must intersect
${\cal C}$.  If an orbit in $\doc$ intersected ${\cal C}$ in more than
one point, there would exist a $q \in \doc$ and a $t > 0$ such that
both $q$ and $\ph_t (q)$ both lie on the boundary of the future of
$p$. Equivalently, $q$ lies on the boundary of the future of both $p$
and $\ph_{-t} (p)$.  But this is impossible, since $p \in I^+(\ph_{-t}
(p))$.

To prove the second statement, we note first that, by arguments similar
to those of the previous paragraph, we may assume without loss of
generality that $p \in {\cal S}$.
By hypothesis, any
generator, $\lambda$, of ${\cal H}$ intersects $\partial {\cal S}$, and, thus,
(since ${\cal S}$ is achronal) contains a point
not lying in $I^+(p)$. On the other hand, since $\lambda$ contains
a point lying in $\partial {\cal S} \subset \pdoc$ and $\lambda$
cannot have a future  endpont, it follows that $\lambda$
enters $I^+(\doc)$. However,
$I^+(\doc) \subset I^+(\alpha_p)$, where
$\alpha_p$ denotes the Killing orbit through $p$. Hence, if we define
$K_t$ to be the subset of generators of ${\cal H}$ which enter
$I^+(\ph_{t} (p))$, we see that $\{ K_t \}$ for $t \in \R$ yields
an open cover of $K$. By compactness of $K$, there exists a
$t_0 \in \R$ such that every generator of ${\cal H}$ enters
$I^+(\ph_{t_0} (p))$. Applying $\ph_{-t_0}$ to this statement, we conclude
that every generator of ${\cal H}$ must enter $I^+ (p)$, and,
hence, $\lambda$ must intersect ${\cal C}$. Finally, to show that
$\lambda$ cannot intersect ${\cal C}$ more than once, we note that any
$q \in {\cal H} \cap {\cal C}$ must lie on a null geodesic in ${\cal C}$,
which, by global hyperbolicity of $\doc$, must have a past endpoint on
$p$. Hence, if $q, r \in \lambda \cap {\cal C}$ with $q \neq r$, one of them
would be connected to $p$ by a future-directed broken null geodesic, and,
thus, could not lie in ${\cal C}$
\hfill$\Box$

Our main result is the following:

\begin{Theorem} \label{T1}
Let $(M,g_{ab})$ be a stationary, asymptotically flat
spacetime containing a single
asymptotically flat region whose domain
of outer communications, $\doc$, is globally hyperbolic.
Suppose that the null energy condition
(\ref{NEC}) holds. Then
\begin{enumerate}\item
 $\leftineq\rh\rightineq $ is simply connected.
 \item
Suppose that there exists an achronal, asymptotically flat slice,
${\cal S}$, of $\doc$, whose boundary in $M$ intersects
the event horizon, ${\cal H}$, of any black holes
in $M$ in a cross-section, $K$.
If $K$ is compact and if ${\cal C'} \setminus {\cal C}_{\ext}$ has compact
closure in M
(where ${\cal C}$, ${\cal C}_\ext$ and ${\cal C'}$ were defined above), then
each connected
component of $K$ is homeomorphic to a sphere.
\end{enumerate}\end{Theorem}

{\bf Remarks:} \begin{enumerate}
\item \label{toporemark} Simple connectedness of $\doc$ is equivalent to simple
connectedness of any Cauchy surface $\Sigma$ for $\doc$. In  particular
it follows from point 1 of Theorem \ref{T1} and of Lemma 4.9 of
\cite{Hempel} that if $\Si$ is  homeomorphic to the interior of a compact
manifold with boundary $\bar\Si$, then each connected component of
$\partial\bar\Si$ is homeomorphic to a sphere.
\item \label{emptyremark}
The set $K$ in point 2 above can be empty --- in that case we obtain
a generalization of the results of \cite{Masoodtopology,BrillLindblom};
{\em cf.\/}  Remark \ref{toporemark} above.
\item Recall that a construction of Carter
 \cite{CarterHI} reduces the question of uniqueness of
 stationary rotating black holes to that of an appropriate harmonic
 map problem. In that construction simple connectedness of
$\leftineq\rh\rightineq $
 plays a key role, compare \cite{Weinstein1,Weinstein2}.
\end{enumerate}

{\bf Proof:} Using the first property of Lemma \ref{L1}, we define a
continuous time function $\tau$ on $\doc$ by the condition that for
each $q \in \doc$, we have $\ph_{-\tau(q)} (q) \in {\cal C}$. It follows
immediately that $\tau$ increases monotonically along any future-directed
timelike curve, and that for all $q \in \doc$ and all $t \in \Bbb R$,
we have $\tau (\ph_t(q)) = \tau(q) + t$. (A smooth time function
on $\doc$ with
these properties could be obtained by the construction of
Proposition 4.1 of \cite{ChW}.)
We thereby obtain the
identification $\leftineq\rh\rightineq
\approx{\R }
\times {\cal C'}$ as already noted in Lemma \ref{L1}.
To prove point 1, it suffices to show that any closed path,
$\ga$, in ${\cal C'}$ is
contractible in ${\cal C'}$. Without loss of generality
we may assume that $p\in\ga$ (where, we recall that
${\cal C} \equiv \partial I^+(p)$).
Without loss of
generality we also may assume that $p$ lies on a two--sphere $S^2\subset
M_{\ext}$ the null inward pointing normals of which are everywhere
converging, where, as defined above,
 $M_{\ext}$ is defined as the orbit of the asymptotically
flat end $\Si_{\ext}$ under the isometries.

Consider first the simpler case where $X^a$ is timelike on
$\leftineq\rh\rightineq $. Let $s\in[0,2\pi]$ be any parameter on
$\gamma$ with $\gamma(0)=\gamma(2\pi)=p$. By compactness of $S^1$ we
can choose a constant $A$ large enough so that the curve $$
[0,2\pi]\ni s\to\Ga(s)=(As,\ga(s))
\in \R\times {\cal C'}
$$ is timelike. The curve $\Ga$ is then a causal curve from
$(0,p)$ to $(2 \pi A,p)$,
so that it follows from
\cite{FriedmanSchleichWitt} that
$\Ga$ is homotopic to the curve $\tilde\Gamma (s)=(As,p)$
keeping both end points fixed. Since ${\cal C'}$ is a deformation retract of
$\R\times{\cal C'}$, contractibility of $\ga$ in ${\cal C'}$ follows.

To cover the case in which ergoregions occur some more work is needed.
Let thus $p$, $\ga$, etc., be as above, and consider any $q\in\ga$. We
first wish to show that there exists a $T\in \Bbb R$ and a future
directed causal curve from $(0,q)$ to $(T,q)$ the projection of which
to ${\cal C'}$ is homotopically trivial. Indeed, since
$(0,q)\in\leftineq\rh\rightineq $, it follows from Lemma 3.1 of \cite{ChW}
that there exists a future directed timelike curve $\Ga_1(s)$,
$s\in[0,1]$, from $(T_1,p)$ to $(0,q)$ for some $T_1\in \Bbb R$.
Similarly there exists a future directed timelike curve
$\Ga_2(s)$, $s\in[0,1]$ from $(0,q)$ to $(T_2,p)$ for some $T_2\in\Bbb R$.
The curve $\Ga_3=\phi_{T_2-T_1}(\Ga_1)\circ \Ga_2$ is then a future
directed timelike curve from $(0,q)$ to $(T_2-T_1,q)$. Note that the
curve  $\Ga_2\circ\Ga_1$ is a causal curve from
$\R\times{\cal C'}_{\ext}$ to
itself, and hence has homotopically trivial projection on ${\cal C'}$ by
\cite{FriedmanSchleichWitt}. But the projections on ${\cal C'}$ of
$\Ga_2\circ\Ga_1$ and of
$\Ga_3$ coincide, which establishes homotopic triviality of $\Ga_3$.

Consider now $q(s)\in\ga$, denote by  $\Ga_s$  the timelike curve from
$(0,q(s))$ to $(T(s),q(s))$ just constructed.
It is convenient to identify $s\in[0,2\pi]$ with some parameter on
$S^1$ in the obvious way. There exists a
neighbourhood ${\cal O}_s\subset{\cal C'}$ of $q(s)$ such that any two
points $r\in\{0\}\times {\cal O}_s$ and $\tilde r\in\{T(s)\}\times
{\cal O}_s$ lie on a causal curve which coincides with $\Ga_s$ except
near its end points. It follows that there exists $\epsilon(s)>0$ such
that for all $s_-\in(s-\epsilon(s),s]$ and $s_+\in[s,s+\epsilon(s))$
there exists a causal curve $\Ga_{s_-,s_+}$ between $(0,\ga(s_-))$ and
$(T(s),\ga(s_+))$. Reducing $\epsilon$ if necessary the projection on
${\cal C'}$ of $ \Ga_{s_-,s_+}$ can be chosen to be homotopic with both ends
fixed to $\ga|_{[s_-,s_+]}$.

Consider finally the covering
$\{(s-\epsilon(s),s+\epsilon(s))\}_{s\in S^1}$ of $S^1$. By compactness
a finite covering $I_i=(s_i-\epsilon(s_i),s_i+\epsilon(s_i))$,
$i=1,\ldots,N$ can be chosen. We can order the intervals $I_i$ in the
obvious way and choose them to be
pairwise disjoint except for the nearest neighbours, with $p\in I_1$
and $p\in I_N$. Let
$\ga(r_0)=p$, $\ga(r_N)=p$, and for $1\le i\le N-1$ let
$r_i$ be any point in $I_i\cap I_{i+1}$. By
construction for all $i$ there exists a causal curve $\Ga_i$ from
$(0,\ga(r_i))$ to $(T(r_i),\ga(r_{i+1}))$. Moreover the projection of
$\Ga_i$ on ${\cal C'}$ is homotopic with both ends fixed to
$\ga|_{[r_i,r_{i+1}]}$. It follows that the  curve
$\phi_{T(r_1)+\ldots+T(r_{N-2})}(\Ga_{N-1})\circ\ldots\circ
\phi_{T(r_0)}(\Ga_1)\circ\Ga_0$ is a causal curve from $\R \times
{\cal C'}_{\ext}$ to itself, the projection $\tilde \ga$ of
which on ${\cal C'}$ is
homotopic to $\ga$. By \cite{FriedmanSchleichWitt} $\tilde\ga$ is
homotopically trivial. It follows that $\ga$ is contractible to $p$,
and point 1 follows.

To prove point 2, we note that the closure in $M$ of
${\cal C'} \setminus {\cal C}_{\ext}$ is a compact manifold with
boundary, whose interior is simply connected. Furthermore, by
Lemma \ref{L1}, the boundary of this manifold is homeomorphic
to the disjoint union of $K$ with a two--sphere in the asymptotic
region. The conclusion then
follows\footnote{We are grateful to
G. Galloway for pointing this out.} directly from
Lemma 4.9 of \cite{Hempel}.
\hfill$\Box$

 {\bf Acknowledgements} P.T.C. wishes to thank G. Galloway for many
useful discussions and for comments on a previous version of this paper.
He is also grateful to the University of Chicago
and to the E. Schr\"odinger Institute in Vienna for hospitality during
work on this paper.
\ifx\undefined\bysame
\newcommand{\bysame}{\leavevmode\hbox to3em{\hrulefill}\,}
\fi

\end{document}